# Improvement of Photophysical Properties of $CsPbBr_3$ and $Mn^{2+}$:$CsPb(Br,Cl)_3$ Perovskite Nanocrystals by $Sr^{2+}$ Doping and Their Application in White-LEDs

Hurriyet Yuce[1], Mukunda Mandal[2], Yenal Yalcinkaya[2], Denis Andrienko[2], Mustafa M. Demir[1*]

[1]Department of Materials Science and Engineering, Izmir Institute of Technology, 35430, Urla, Izmir, Turkey

[2]Max Planck Institute for Polymer Research, Ackermannweg 10, 55128 Mainz, Germany

*Corresponding Author: Mustafa M. Demir, mdemir@iyte.edu.tr

**ORCID:**

Hurriyet Yuce 0000-0002-7459-2833
Mukunda Mandal: 0000-0002-5984-465X
Denis Andrienko 0000-0002-1541-1377
Mustafa M. Demir 0000-0003-1309-3990

**Abstract**

All-inorganic lead halide perovskite nanocrystals (NCs) show potential in optoelectronic devices, though their stability and efficiency have yet to improve. In this work, we explore the effect of bivalent metal site doping on the optoelectronic properties of $CsPbX_3$ (X = Br, Cl) perovskite NCs. First, the $Pb^{2+}$ ions in pristine $CsPbBr_3$ NC are partially substituted by $Mn^{2+}$ ions, and the alkaline earth metal strontium is then doped on both pristine and $Mn^{2+}$-substituted particles. The structural and photophysical properties of the pristine and the three doped NC variants have been investigated experimentally as well as exploiting first principles calculations. We found that a small percentage of $Sr^{2+}$-doping improved the photoluminescence quantum yield of the pristine particle by 8%, while it improved the $Mn^{2+}$-state emission by 7%. Perovskite NC film/poly(methyl methacrylate) composites with all four NC variants were used in a white light-emitting diode (WLED) and again both $Sr^{2+}$ doped NCs were found to increase the luminous efficiency of the WLED by ca 4 %. We attribute this performance enhancement to a reduced defect density as well as an attenuated microstrain in the local NC structure.

## Introduction

All-inorganic lead halide perovskite nanocrystals (NCs), e.g., CsPbX$_3$ (X = Cl$^-$, Br$^-$, I$^-$) have been used to manufacture LEDs,[1–3] solar cells,[4,5] lasers,[6,7] and photodetectors.[8,9] High photoluminescence quantum yield (PLQY), defect tolerance, adjustable bandgaps, narrow photoluminescence (PL), and low cost are assets of perovskite NCs[10–12]. The issues preventing their use in real-world applications are their stability and surface defects, which promote non-radiative recombination and therefore reduce PLQY.[13,14] Surface passivation by polymers or inorganic ligands[15,16] reduces the defect concentration, but often leads to poor conductivities of NC films. Alternatively, compositional engineering can be used to increase stability and eliminate defects,[14,17,18] either with homovalent dopants, such as Zn$^{2+}$[19,20], Cd$^{2+}$[19,21], Mn$^{2+}$[22,23], and Sn$^{2+}$[19,24] or heterovalent dopants, such as Ce$^{3+}$[25], Bi$^{3+}$[26], Sb$^{3+}$[27], and Al$^{3+}$[28]. For example, Sr$^{2+}$ homovalent dopants for CsPbI$_3$ perovskite NCs improve their stability and optical properties by increasing defect formation energy.[29,30] Mn$^{2+}$ incorporation into the perovskite lattice results in a mid-gap Mn$^{2+}$ states in the band structure, which affects the recombination dynamics[31]. Mn-doping results in a PL emission just below the bandgap (~2.4 eV for CsPbBr$_3$ or ~3.0 eV for CsPbCl$_3$), assigned to exciton recombination in the perovskite, as well as emission around 2.06 eV[32], assigned to the Mn $d$-$d$ transition between the $^4$T1 and $^6$A1 configurations[31,33].

For perovskite NC-based white light emitting diodes (WLEDs), CsPbX$_3$ NCs can be employed to obtain white light with the combination of the emissions in red-green-blue regions[3,34]. To achieve red PL emission, Mn-doped perovskite NCs can be used as an alternative to CsPbI$_3$ NCs, since letter suffer from thermodynamic instability at room temperature[10]. CsPbBr$_3$ exhibits better phase stability compared to CsPbI$_3$ NCs[34]. A low PLQY of Mn state emission is still a problem for the commercial usage. However, Bi$^{3+}$[35], Co$^{2+}$[36], Ni$^{2+}$[37], Yb$^{3+}$[38], and Cu$^{2+}$[39] co-dopants can be used to improve the PLQY of Mn state emission in Mn$^{2+}$-doped perovskite NCs.

We report the improved photophysical properties of CsPbBr$_3$ and Mn$^{2+}$:CsPb(Br,Cl)$_3$ perovskite NCs achieved by Sr$^{2+}$ doping, with the highest PLQY observed for 2 % of Sr$^{2+}$ dopant. To harness emission in the red region, the 2% Sr$^{2+}$ sample was further doped with MnCl$_2$ yielding Mn$^{2+}$:CsPb(Br,Cl)$_3$ NCs. For both Sr$^{2+}$:CsPbBr$_3$ and Mn$^{2+}$:CsPb(Br,Cl)$_3$ NCs we observed improved PLQY and PL lifetimes, leading to a superior performance of these NCs in a white LED.

## Results and discussion

**Experimental Results**

We carried out ICP-MS measurements (Table S4 and S5) in order to determine actual doping ratio of Sr/Pb for $CsPbBr_3$ NCs and the ratios were estimated as 0.047%, 0.099%, 0.238%, 1.051% for the 1%, 2%, 5%, and 10% $Sr^{2+}$ concentration, respectively. For $Mn^{2+}$:$CsPb(Br,Cl)_3$ perovskite NCs, actual doping ratio of 2% $Sr^{2+}$ was estimated as 0.0205%. ICP-MS analysis results show that $Sr^{2+}$ was successfully doped into perovskite NCs and doping concentrations of $Sr^{2+}$ against their addition amounts are in agreement with $Sr^{2+}$:$CsPbI_3$ NCs[29]. $Br^-$/$Cl^-$ ratios were determined as 72.34% and 75.26% for $Mn^{2+}$:$CsPb(Br,Cl)_3$ and 2% $Sr^{2+}$: $Mn^{2+}$:$CsPb(Br,Cl)_3$ by IC analysis. Large difference in bond dissociation energies between ions causes to prevent the formation of doping into perovskite NCs[22]. Considering the bond dissociation energies of Sr-Br (365 kJ/mol) and Sr-Cl (409 kJ/mol),[40] incorporated $Sr^{2+}$ within lattice could be favoring doping of $Br^-$ compared to the $Cl^-$.

The morphology and size distribution of $CsPbBr_3$ NCs upon $Sr^{2+}$ doping was monitored via STEM mode in SEM. Incorporation of $Sr^{2+}$ ions to the perovskite lattice did not clearly affect the structure and morphology (Fig.1a). From the STEM images, the average NC sizes of pristine, 1%, 2%, 5%, and 10% $Sr^{2+}$-doped $CsPbBr_3$ NCs were found to be ~20.3, ~16.8, ~17.3, ~16.0, and ~17.0 nm, respectively. While a systematic reduction of perovskite NC size corresponding to an increasing $Sr^{2+}$ ion concentration was not observed, in general however, the $CsPbBr_3$ NCs experienced a reduction in nanoparticle size due to $Sr^{2+}$ doping. Passivation by excess halide ions have been reported to cause a lower size distribution in perovskite NCs[41,42], however, we hypothesize that the reason for the decrease in NC size upon $Sr^{2+}$ doping (Figure 1) is mainly an effect of impurity (i.e., the dopant) on crystallization dynamics, since the synthesized perovskite NCs are stoichiometric in this work. We argue that the $Sr^{2+}$ impurities lead to a reduction of the energy threshold for nucleation and cause more nuclei to form, resulting in a smaller NC size. The uniformity of NC size distribution is improved upon $Sr^{2+}$ doping until 2% $Sr^{2+}$ concentration, however, even higher $Sr^{2+}$ concentrations adversely affect NC homogeneity (Figure 1(a-e) insets).

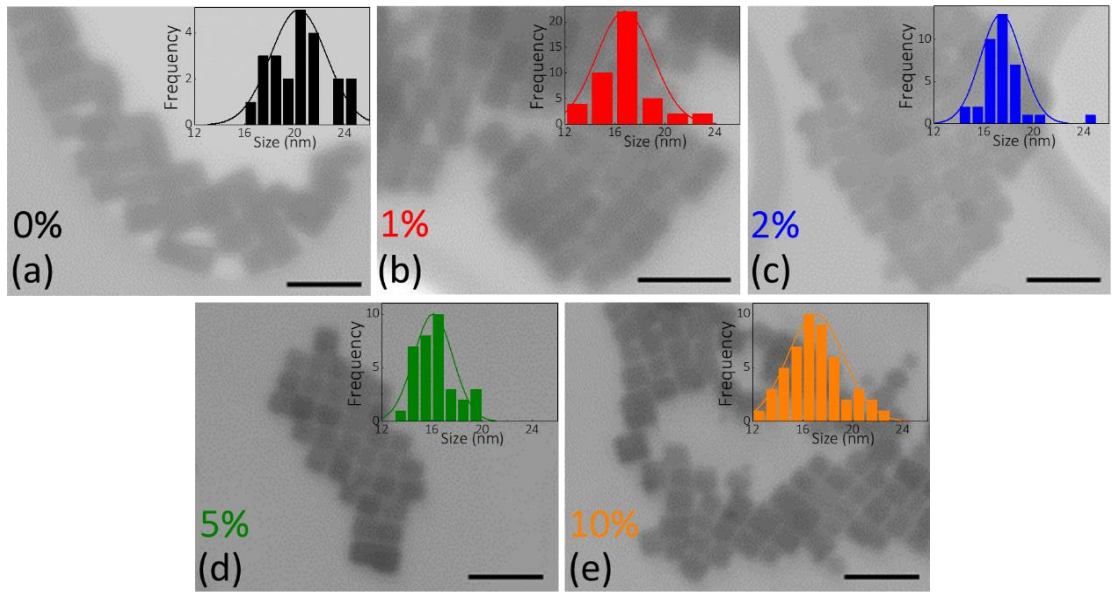

Figure 1. STEM images of (a) 0%, (b) 1%, (c) 2%, (d) 5%, and (e) 10% $Sr^{2+}$-doped $CsPbBr_3$ NCs. All inset in each figure shows NC size distribution in relevant sample (Scale of x-axis for all figure are kept constant.). Bar size is 50 nm.

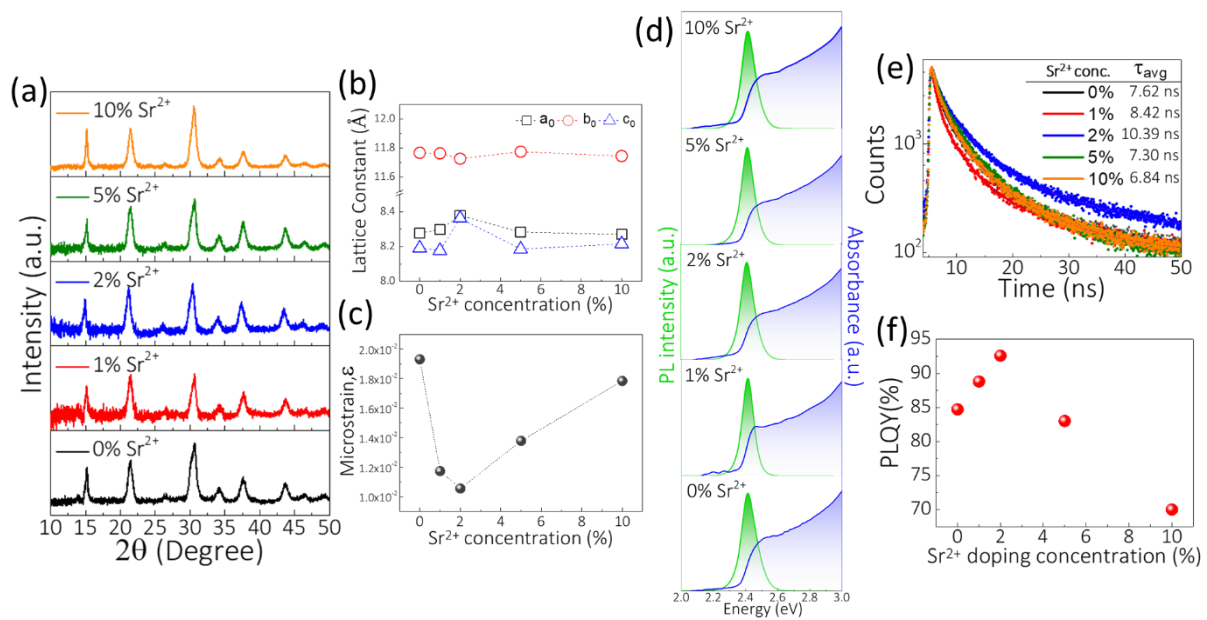

Figure 2. (a) X-ray diffraction patterns, (b) lattice parameters, (c) microstrain values obtained by W-H plot, (c) PL/Absorption spectra, (d) TRPL plot, and (e) PLQY distribution of pristine and $Sr^{2+}$:$CsPbBr_3$ NCs.

Figure 2(a) shows the XRD patterns of pristine and $Sr^{2+}$:$CsPbBr_3$ NC samples. XRD reflections for all samples confirm the orthorhombic Pnma crystal structure of $CsPbBr_3$[43]. Incorporation of $Sr^{2+}$ to perovskite lattice does not cause any additional XRD reflection, confirming no phase transition upon doping. Furthermore, no systematic shift in XRD reflections was observed upon Sr$^{2+}$ doping due to the similar ionic radii of Pb$^{2+}$ (119 pm) and Sr$^{2+}$ (118 pm) ions[44]. In order to understand the effect of Sr$^{2+}$ doping on the perovskite lattice, the lattice parameters were calculated by using XRD patterns of each sample as given in Table S1. The lattice parameters (a, b, c unit cell) show a trend where $a$ = 8.28 Å and "c= 8.19 Å values increase and reach to a maximum at 2% Sr$^{2+}$ doping concentration and then decrease as the doping concentration exceeds 2% (Figure 2(b)) whereas $b$ lattice parameter follows the opposite trend. The absence of a systematic shift in XRD patterns (i.e., homogenous strain) has led to investigation of microstrain depending on the doping concentration. Microstrain describes local distortion of the crystal lattice, which was calculated by the slope of Williamson-Hall (W-H) plot obtained by the XRD reflection broadness of the samples[45,46]. According to the calculations in Figure 2(c), microstrain has declined upon Sr$^{2+}$ doping until 2% concentration. For further concentrations, there is a systematic increase in microstrain. This behaviour suggests reduced defect related distortion in the perovskite NCs at 2% Sr$^{2+}$ could be the result of changed lattice parameters upon doping.

Table 2. PLQY and PL lifetime of various NC samples having different %Sr-doping.

| Sr$^{2+}$:CsPbBr$_3$ System | PLQY | PL Lifetime [ns] |
|---|---|---|
| 0% Sr$^{2+}$ (pristine) | 84.7% | 7.62 |
| 1% Sr$^{2+}$ | 88.8% | 8.42 |
| 2% Sr$^{2+}$ | 92.6% | 10.39 |
| 5% Sr$^{2+}$ | 83.0% | 7.30 |
| 10% Sr$^{2+}$ | 70.0% | 6.84 |

In order to understand the doping effect further, we have carried out optical measurements as shown in Figure 2(d-f). Considering absorption and PL spectra of pristine and Sr$^{2+}$:CsPbBr$_3$ NCs (Figure 2d), there is no shift in PL emission energy and absorption edge of the samples. This implies that the incorporation of Sr$^{2+}$ to the lattice does not affect the bandgap energy of CsPbBr$_3$. According to the Tauc's plot (Figure S2) the bandgap energies of the samples are calculated as ∼ 2.39 eV. PL signals were centered at 2.41 eV. Furthermore, the full-width half maxima (FWHM) of PL peaks (Figure S3) were reduced by the doping of Sr$^{2+}$ which is in an agreement with size distribution in STEM images. The absence of PL shift despite the minor NC size changes can be attributed to sizes of CsPbBr$_3$ NCs within intermediate-weak confinement regimes[47]. Figure 2(e) shows the TRPL results of CsPbBr$_3$ NCs with different Sr$^{2+}$ doping amounts. The average PL lifetimes per sample were calculated using equation S4. PL lifetime of CsPbBr$_3$ NCs was measured as 7.62 ns. In case of the incorporation of Sr$^{2+}$ ions until ≤ 2% to the perovskite NCs, we observed longer average lifetimes. Depending on the doping concentration, there is a systematic increase in average lifetime which are 8.42 and 10.39 ns for 1% and 2% Sr$^{2+}$-doped CsPbBr$_3$ NCs, respectively. However, for higher concentrations of Sr$^{2+}$ (>2%), we found a reduction of lifetimes which are 7.30 ns for 5% and 6.84 ns for 10% Sr$^{2+}$ perovskite NCs. Considering inversely proportional defect-lifetime relationship[48,49], the trend in PL lifetimes is in an agreement with calculated microstrain results in Figure 2(c). According to the microstrain calculations, defect-related distortions were diminished by incorporation of Sr$^{2+}$ into the lattice until 2% concentration and minimum microstrain originating from fewer defects was determined for 2% Sr$^{2+}$-doped CsPbBr$_3$ NCs. Then, higher doping amount of Sr$^{2+}$, a reduction of PL lifetime was observed with increasing microstrain due to higher defect concentration resulting in more non-radiative recombination. The reduced defect density leads to decreased non-radiative recombination, resulting in higher PLQY and PL lifetime[50]. We observed the same trend as in PL lifetime and microstrain for PLQY results in Figure 2(f). The PLQY of CsPbBr$_3$ NCs was obtained as 84.7%. By doping of Sr$^{2+}$, PLQY of the perovskite NCs was enhanced to 88.8% and 92.6% at 1% and 2% Sr$^{2+}$ concentrations, respectively. For further doping concentrations, PLQY of the perovskite dropped to 83% at 5% Sr$^{2+}$ and then 70% at 10% Sr$^{2+}$ concentrations since more dopants in the perovskite lattice could trigger the formation of new defects.[30]

Figure 3 shows the STEM images of Mn$^{2+}$:CsPb(Br,Cl)$_3$ and 2% Sr$^{2+}$:MnCl$_2$:CsPbBr$_3$ NCs. 2% Sr$^{2+}$ incorporation into Mn$^{2+}$:CsPb(Br,Cl)$_3$ lattice is benign to the NC morphology shape-wise. However, a slight decrease in nanocube size and improved monodispersity in size were observed similar to the Sr$^{2+}$:CsPbBr$_3$ case (Figure 1). The average NC sizes are 20.6 nm and 19.3 nm for pristine and 2% Sr$^{2+}$:Mn$^{2+}$:CsPb(Br,Cl)$_3$, respectively.

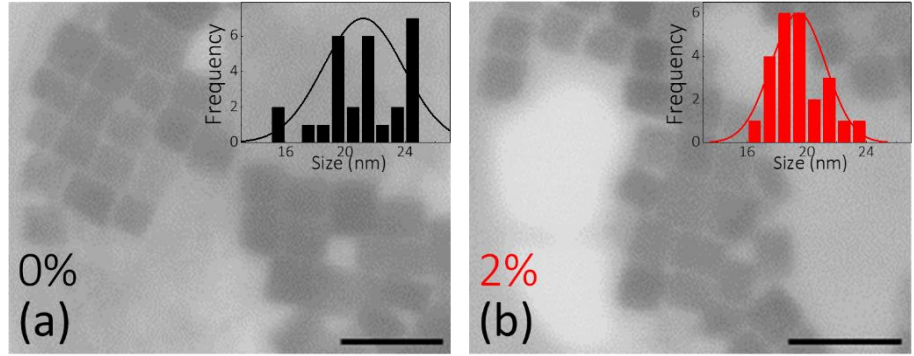

Figure 3. STEM images of (a) $Mn^{2+}$:CsPb(Br,Cl)$_3$ and (b) 2% $Sr^{2+}$:$Mn^{2+}$:CsPb(Br,Cl)$_3$ NCs. Bar size is 50 nm.

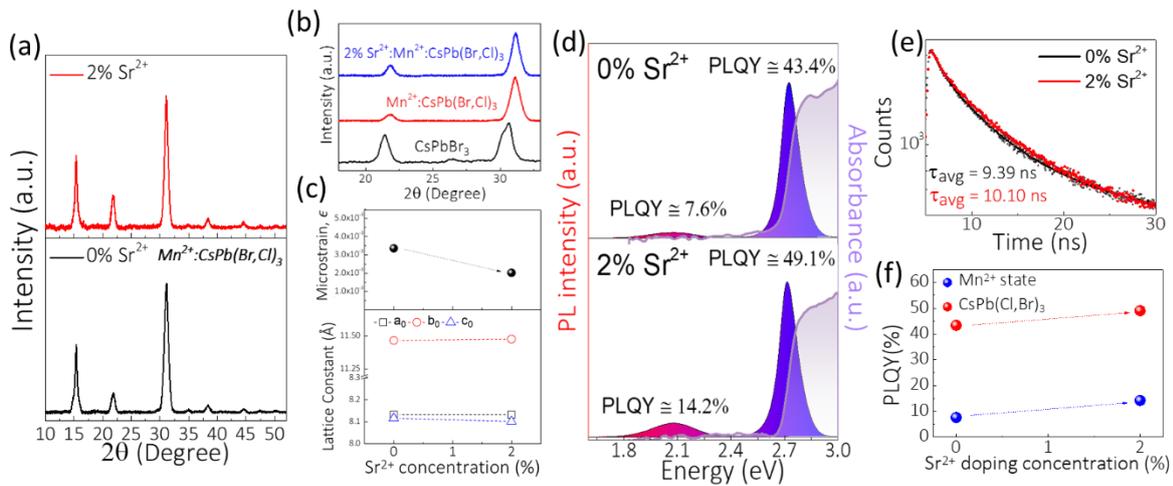

Figure 4. (a) X-ray diffraction patterns, (c) microstrain and lattice constant parameters, (d) PL/Absorption spectra, (e) TRPL plot, and (f) PLQY comparisons of $Mn^{2+}$:CsPb(Br,Cl)$_3$ and 2% $Sr^{2:}$ $Mn^{2+}$:CsPb(Br,Cl)$_3$ NCs. (b) The XRD signal position comparisons of CsPbBr$_3$ and $Mn^{2+}$:CsPb(Br,Cl)$_3$ perovskite NCs.

XRD analyses for $Mn^{2+}$:CsPb(Br,Cl)$_3$ NCs in order to observe the impact of 2% $Sr^{2+}$ doping on the NCs were carried out as show in Figure 4(a). The XRD pattern of $Mn^{2+}$:CsPb(Br,Cl)$_3$ NCs show reflections at 15.4º, 22.0º, 31.1º, 35.0º, 38.4º, and 44.6º which correspond to the orthorhombic perovskite phase[51]. Due to ionic radius difference between $Mn^{2+}$ (67 pm) and $Pb^{2+}$ (119 pm) ions[44], the XRD signal shift to higher angles for $Mn^{2+}$-doped perovskite NCs, which indicates $Mn^{2+}$ ions were incorporated into the perovskite lattice[52]. Nonetheless, the shift in $Mn^{2+}$:CsPb(Br,Cl)$_3$ is not exclusively due to $Pb^{2+}$ ion substitutions with $Mn^{2+}$ ions, the overall lattice shrinking effect should also be attributable to $Cl^-$/$Br^-$ substitutions Figure 4(b)[34]. $Sr^{2+}$ addition into $Mn^{2+}$:CsPb(Br,Cl)$_3$ NCs does not exhibit any peak shift, indicating the $Sr^{2+}$ ions substituted with $Pb^{2+}$ ions in the perovskite NCs. Therefore, we have performed microstrain calculations of $Mn^{2+}$:CsPb(Br,Cl)$_3$ samples owing to the narrowing of XRD signals after $Sr^{2+}$ incorporated. By the doping, microstrain of $Mn^{2+}$:CsPb(Br,Cl)$_3$ NCs diminishes from 0.33% to 0.20% as shown in Figure 4(c). In the lattice parameters, b and c exhibit an inversely proportional trend while "a=8.1324 Å" and stays constant by $Sr^{2+}$ doping (Figure 4(c) and Table S1). The UV-Vis results in Figure 4(d) show very similar absorption behaviour for $Mn^{2+}$:CsPb(Br,Cl)$_3$ and 2%$Sr^{2+}$:$Mn^{2+}$:CsPb(Br,Cl)$_3$ NCs. The Tauc plots shown in Figure S4 give 2.70 and 2.69 eV bandgaps for these samples, respectively. This small variation between the samples was also observed in PL peak positions obtained from excitons where $Mn^{2+}$:CsPb(Br,Cl)$_3$ and 2%$Sr^{2+}$ $Mn^{2+}$:CsPb(Br,Cl)$_3$ NCs exhibited exciton PL signals at 2.72 and 2.71 eV, respectively (Figure 4(d)). Since we have shown that $Sr^{2+}$ doping at low concentrations ($\leq$ 1%) does not cause a change in the bandgap energy (Figure S2), this slight change in bandgap and PL emission energies can be attributed to extremely small variety of $Br^-$/$Cl^-$ ratio in the perovskite structures. PL emission coming from Mn-state in $Mn^{2+}$:CsPb(Br,Cl)$_3$ perovskite NCs increases by $Sr^{2+}$ doping. TRPL measurement results of these samples are given in Figure 4(e) for exciton emissions of the perovskites. The average PL lifetime of $Mn^{2+}$:CsPb(Br,Cl)$_3$ NCs was increased from 9.39 ns to 10.10 ns by $Sr^{2+}$ doping, indicating lower defect density. As the last optical measurement, we have performed PLQY measurements on $Mn^{2+}$:CsPb(Br,Cl)$_3$ and 2%$Sr^{2+}$:$Mn^{2+}$:CsPb(Br,Cl)$_3$ NCs. For these measurements, only the range between 2.38-1.59 eV (520-780 nm) and 3.02-2.38 eV (410-520 nm) were included to see the change in the optical performance of $Mn^{2+}$-state and perovskite exciton emissions, respectively. As a result, $Mn^{2+}$:CsPb(Br,Cl)$_3$ NCs have a PLQY of 7.6% and they show a remarkable increase in the PLQY reaching up to 14.2% after $Sr^{2+}$ incorporation for $Sr^{2+}$:$Mn^{2+}$:CsPb(Br,Cl)$_3$ NCs as given in Figure 4(f), proving the beneficial effect of $Sr^{2+}$ co-doping alongside with $Mn^{2+}$. In addition, the exciton PLQY belonging to the perovskite structure is increased from 43.4% to 49.1% by $Sr^{2+}$ incorporation as well (Figure 4(f)). These optical improvements with better exciton-to-$Mn^{2+}$ energy transfer due to $Sr^{2+}$ incorporation are originated from lowered defect density.

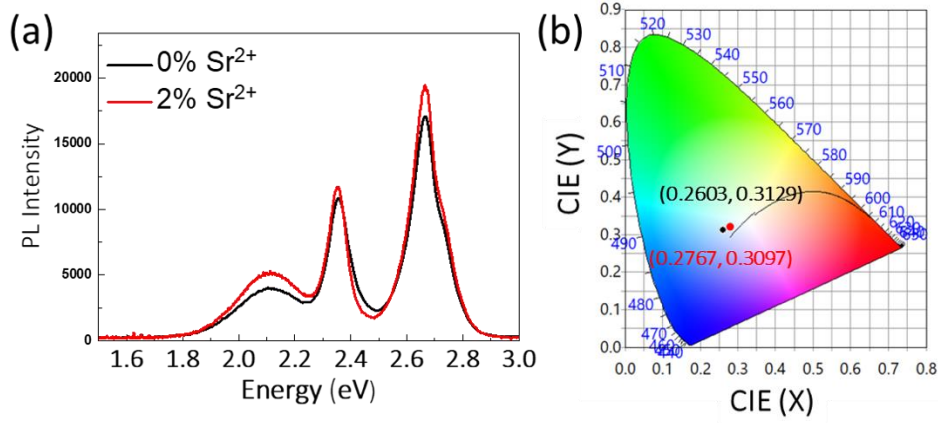

Figure 5. PL spectra of WLEDs based on (a) 0% and 2% $Sr^{2+}$ doped $CsPbBr_3$ and $Mn^{2+}$:$CsPb(Br,Cl)_3$ dropcasted films; (b) corresponding CIE coordinates.

Two sets of WLEDs were fabricated based on $CsPbBr_3$/PMMA (green emitting composite) and $Mn^{2+}$:$CsPb(Br,Cl)_3$/PMMA (red-blue emitting composite) by excitation of an UV LED. In order to observe 2% $Sr^{2+}$ addition effect on the quality of WLED, perovskite NCs/PMMA composite films with and without 2% $Sr^{2+}$ were obtained by drop casting. Their corresponding PL spectra for 0% $Sr^{2+}$ and 2% $Sr^{2+}$ systems are shown in Figure 5(a) with their CIE coordinates in Figure 5(b). Since the amount of $Mn^{2+}$:$CsPb(Br,Cl)_3$ perovskite NCs in the film is higher compared to the amount of $CsPbBr_3$ NCs, PL emissions of $Mn^{2+}$:$CsPb(Br,Cl)_3$ is improved more by $Sr^{2+}$ doping compared to the improvement of $CsPbBr_3$ NCs. The luminescence efficiency (LER) of $CsPbBr_3$/$Mn^{2+}$:$CsPb(Br,Cl)_3$ based WLED is 277 lm/W under the 8W power of UV light, which is increased to 290 lm/W by $Sr^{2+}$ doping. While the CRI for the undoped perovskite system is 80, 2% $Sr^{2+}$:$CsPbBr_3$/2% $Sr^{2+}$:$Mn^{2+}$:$CsPb(Br,Cl)_3$ based WLED exhibit a CRI of 83 due to the increase in PL emission of Mn- state by decreasing the deficiency of red emission in the system.

**Computational Results**

In order to estimate the relative propensity for the formation of a given nanocrystal (NC), we first constructed three cubic NCs of size ~3 nm. Their corresponding cohesive energy ($E_{cohesive}$), which quantifies the energy required to dissociate the atoms constituting the NC into a collection of neutral free atoms, was then calculated using equation (1)

$$E_{\text{cohesive}} = \frac{\sum_i n_{\text{atom},i} E_{\text{atom},i} - E_{NC}}{n_{NC}} \tag{1}$$

where $n_{atom,i}$ is the total number of atoms of type $i$ constituting the NC, $E_{atom,i}$ is its corresponding isolated neutral ground state energy, $E_{NC}$ is the total energy of the NC in its optimized geometry and $n_{NC}$ is the total number of atoms forming the NC. For the pristine, $Sr^{2+}$-doped and $Sr^{2+}/Mn^{2+}/Cl^-$-doped systems, the $E_{cohesive}$ was computed to be 3.04, 3.05 and 3.17 eV/atom respectively. Thus, while $Sr^{2+}$ doping replacing a $Pb^{2+}$ ion is favorable only slightly ($\Delta E_{cohesive}$ ~ 0.01 eV/atom), the $Cl^-$-anions from $MnCl_2$ drives the formation of $Sr^{2+}:CsPb(Br,Cl)_3$, $Mn^{2+}$ having $\Delta E_{cohesive}$ ~ 0.13 eV/atom with respect to the pristine NC.

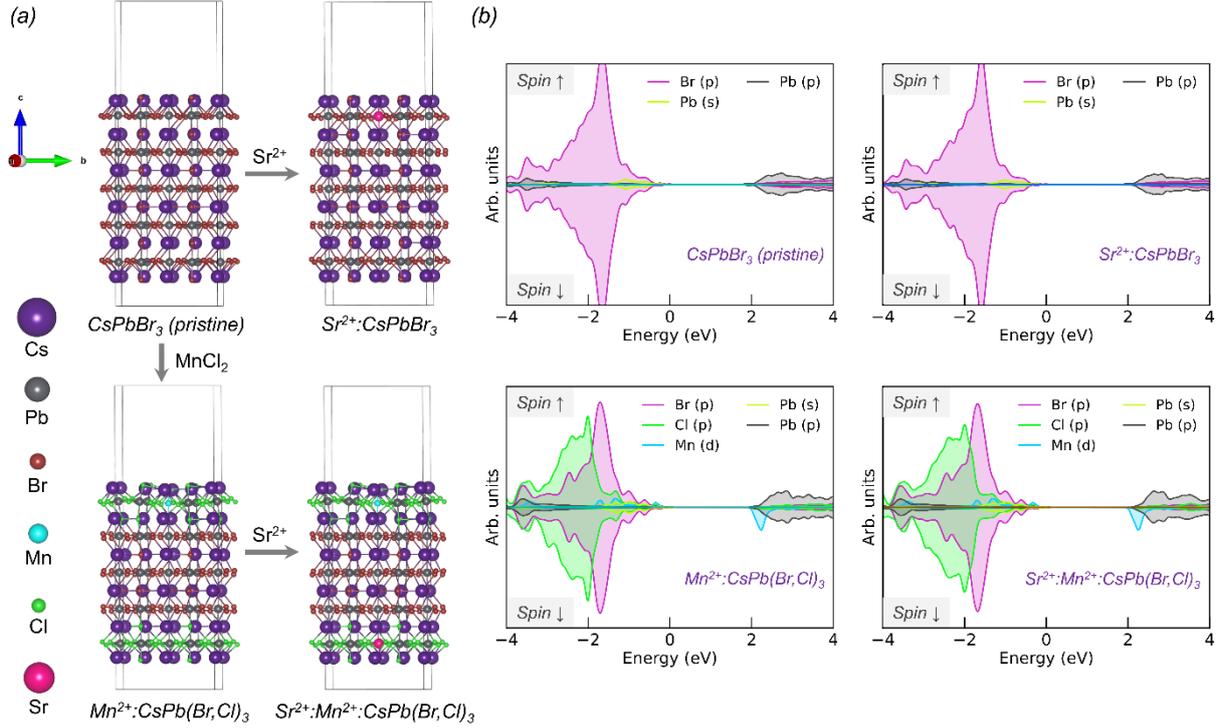

**Figure 6.** Four periodic models considered for this study (a): Pristine $CsPbBr_3$, $Sr^{2+}$-doped $CsPbBr_3$, $Mn^{2+}$-doped $CsPbBr_3$ and $Sr^{2+}$ and $Mn^{2+}$ co-doped $CsPb(Br,Cl)_3$. (b) Their corresponding projected density of states.

Fig. 6a describes the four slab models studied in this work: the pristine $CsPbBr_3$ particle, 5% Sr-doped particle $Sr^{2+}:CsPbBr_3$, 5% Mn-doped particle $Mn^{2+}:CsPb(Br,Cl)_3$, and a Sr/Mn-co-doped particle $Sr^{2+}:Mn^{2+}:CsPb(Br,Cl)_3$. The orbital projected density of states of all four models are demonstrated in Fig. 6b, and their band structures are shown in the Supporting Information.

While the density of states of pristine NCs are not very much affected by doping, except expected sub gap states due to $Mn^{2+}$ (see Figure 6 for details), the workfunction is significantly reduced upon doping, see **Table 3**. The direct implication is a smaller charge injection barrier, and hence PLQY and WLED efficiency, as indeed reported experimentally.

**Table 3.** Computed work function and band gaps of four NCs considered in this study.

| NC Particle | Work function (eV) | Band gap (eV) |
| --- | --- | --- |
| CsPbBr$_3$ | 4.43 | 1.89 |
| Sr$^{2+}$:CsPbBr$_3$ | 4.38 | 1.98 |
| Mn$^{2+}$:CsPb(Br,Cl)$_3$ | 4.28 | 2.06 |
| Sr$^{2+}$:Mn$^{2+}$:CsPb(Br,Cl)$_3$ | 4.19 | 2.11 |

## Conclusions

We have presented detailed experimental and computational studies on Sr$^{2+}$ doping into CsPbBr$_3$ and Mn$^{2+}$:CsPb(Br,Cl)$_3$ perovskite NCs. Upon Sr$^{2+}$ doping, a decrease in the NC size of the perovskite was observed by STEM images. The changes in lattice parameters and microstrains were calculated from XRD signals, where microstrain reaches a minimum at 2% Sr$^{2+}$ concentration for CsPbBr$_3$ NCs possibly due to elimination of defects. In addition, the improved optical properties with a PLQY of 92.6% increased from 84.7% by reducing non-radiative recombination with fewer defects. Based on the results obtained by Sr$^{2+}$:CsPbBr$_3$ NCs, we have employed the same strategy to improve the optical properties of Mn$^{2+}$:CsPb(Br,Cl)$_3$ NCs, which was characterized by the same techniques. The PLQY of Mn$^{2+}$-state emission is increased to 14.2% from 7.6%. Finally, improved 2% Sr$^{2+}$-doped CsPbBr$_3$ and Mn$^{2+}$:CsPb(Br,Cl)$_3$ NCs were used to build a WLED resulting in improved white light compared to the WLED build with pristine perovskite NC counterparts. We show that Sr$^{2+}$ doping is a very effective method to make CsPbBr$_3$ and Mn$^{2+}$:CsPb(Br,Cl)$_3$ perovskite NCs more optically active.

## Acknowledgements

This work was supported by the DFG under grant AN680/6-1 (project no. 424708673).

## Experimental Procedure

**Materials**

Lead(II) bromide ($PbBr_2$, ≥98%), strontium bromide ($SrBr_2$, 99.99% trace metal basis), Manganese(II) chloride tetrahydrate ($MnCl_2 \cdot 4H_2O$, ≥98%) cesium carbonate ($Cs_2CO_3$, 99.9%, Sigma-Aldrich), 1- octadecene (ODE, 90%), oleylamine (OLAM, 70%), and oleic acid (OA, 90%) were purchased from SigmaAldrich. Toluene (≥99%, Merck) was purchased and used without any further purification.

**Synthesis of Cs-Oleate for $CsPbBr_3$ NCs**

203.5 mg (266.67 mg) $Cs_2CO_3$, 625 µL (1,167 µL) oleic acid, and 10 mL 1-octadecene were put into a round bottom flask and dried at 120 °C while stirring for 1 h. After drying the temperature was increased to 150 °C under a nitrogen atmosphere. The solution was kept at 150 °C overnight to make sure all $Cs_2CO_3$ reacted with oleic acid. The values given in parentheses are for Cs-Oleate synthesis of pristine and $Sr^{2+}:Mn^{2+}:CsPb(Br,Cl)_3$ NCs.

**Synthesis of pristine and $Sr^{2+}:CsPbBr_3$ NCs**

For the synthesis of pristine $CsPbBr_3$ NCs, first 17.2 mg $PbBr_2$ was put into a glass tube alongside with 1.25 mL ODE, 125 µL OA, and 125 µL OLAM and dissolved/degassed at 120 °C for 30 minutes. After degassing, the temperature was increased to 180 °C where 100 µL Cs-oleate was injected. After injection, the solution was immediately cooled down using an ice bath and then centrifuged for 15 min at 6000 rpm. For the synthesis of $Sr^{2+}:CsPbBr_3$ NCs, same procedure was followed except 1%, 2%, 5% and 10% $SrBr_2$ was added to the solution while the same molar amount of $PbBr_2$ was excluded.

**Synthesis of $Mn^{2+}:CsPb(Br,Cl)_3$ and $Sr^{2+}:Mn^{2+}:CsPb(Br,Cl)_3$ NCs**

For the synthesis of $Mn^{2+}:CsPb(Br,Cl)_3$ NCs, firstly 9.175 mg $PbBr_2$ and 4.95 mg $MnCl_2.4H_2O$ were put into a glass tube alongside with 1.25 mL ODE, 125 µL OA, and 125 µL OLAM and dissolved/degassed at 120 °C for 30 minutes. After degassing, the temperature was increased to 180 °C where 100 µL Cs-oleate was injected. After injection, the solution was immediately cooled down using an ice bath and then centrifuged for 15 min at 6000 rpm. For the synthesis of $Sr^{2+}:Mn^{2+}:CsPb(Br,Cl)_3$ NCs, same procedure was followed except 2% $SrBr_2$ was added to the solution while the same molar amount of $PbBr_2$ was excluded.

**Fabrication of WLED**

0.5 g poly(methyl methacrylate) (PMMA) was dissolved in 10 mL Toluene. For each film of $CsPbBr_3$ drop cast films were prepared with the mixture of 40 μL PMMA and 120 μL perovskite while $Mn^{2+}$:$CsPb(Br,Cl)_3$ drop cast films were prepared with the mixture of 40 μL PMMA and 240 μL perovskite solution on 1.5cm x 1.5cm glass substrates . Then, these $CsPbBr_3$ and $Mn^{2+}$:$CsPb(Br,Cl)_3$ (for 0% and 2% $Sr^{2+}$ series separately) were placed on the LED laterally. The emission spectra were taken from the perovskite layers excited with UV light (366 nm of wavelength, 12 V, and 8W power) and a CIE chromaticity diagram was extracted from the OSRAM ColorCalculator program.

**Characterization**

STEM images for all perovskite NCs were taken by using scanning electron microscopy (SEM); Quanta 250, FEI, Hillsboro, OR. Perovskite solutions were dropped onto 300 mesh holey cabon-Cu (50 micron) for STEM images. X-ray diffraction (XRD) analysis were done by using X'Pert Pro, Philips, Eindhoven, the Netherlands. Optical measurements including absorption, PL, PL lifetime and PLQY were carried out on a FS5 spectrofluorometer (Edinburgh Instruments, U.K.). For PL and PLQY measurements, excitation wavelengths of $CsPbBr_3$ and $Mn^{2+}$:$CsPb(Br,Cl)_3$ samples were 400 nm and 390 nm, respectively. For lifetime measurements, $CsPbBr_3$ and $Mn^{2+}$:$CsPb(Br,Cl)_3$ samples were excited with a 442 nm and 307 nm lasers, respectively. Trace metal analysis was carried out using inductively Coupled Plasma-Mass Spectrometer (ICP-MS) on Agilent 7500ce Octopole Reaction System.

**Computational Methodology**

Periodic calculations were performed using a plane wave basis set implementation of density functional theory within the Vienna Ab initio Simulation Package (VASP, version 6.1) employing the PBE exchange-correlation functional. During all calculations, van der Waals interactions were incorporated employing Grimme's D3 method. The valence-core interactions were described with the projected augmented wave (PAW) method. A plane-wave energy cutoff of 400 eV was used in all calculations. Forces of each atom smaller than 0.02 eV/Å were used during geometry relaxation. The structural relaxation was done by sampling the Brillouin zone over a 3×3×1 k-point grid centered at the Γ point. Slab models were constructed from orthorhombic $CsPbX_3$ (X = Br, Cl) structure considering a 1×2×3 supercell and exposing the CsX terminated surface. Consecutive slabs have been separated by a vacuum of approximately

20 Å to ensure decoupling with periodic image. Visualization and post-processing of the projected density of states plots were performed using the VESTA and sumo packages.

Computations on the non-periodic cluster models are performed using the CP2K 8.1 program suite employing the PBE exchange correlation functional, MOLOPT DZVP basis-set and GTH pseudopotentials for core electrons. A dual basis of localized Gaussians and plane waves (GPW) with a 350 Ry plane-wave cutoff are used for all calculations. Grimme's DFT-D3 protocol was used to account for van der Waals (VDW) interaction. SCF convergence criterion was set at $10^{-6}$ for all calculations.

Initial geometries of CsPbBr$_3$ NCs were obtained by cutting small cubes (~2.8 nm) from the bulk, exposing the CsX layer (X = Br, Cl) at the surface and maintaining overall charge neutrality of the particle. All calculations invoked a nonperiodic boundary condition where, nanoparticles were placed inside a large box of size 50×50×50 Å$^3$ ensuring large vacuum layer above the surface of the NC. All structures were then optimized in vacuum using the BFGS optimizer, setting a maximum force of 5 meV Å$^{-1}$ (1.0×10$^{-04}$ Hartree/Bohr) as convergence criteria.